\documentclass[a4paper,10pt]{article}
\usepackage{enumerate}
\usepackage{color}
\usepackage[utf8]{inputenc} 
\usepackage[english]{babel}
\usepackage[T1]{fontenc}
\usepackage{graphicx}
\usepackage{amsfonts,amssymb,amsmath,latexsym,amsthm}
\usepackage{textcomp}
\usepackage[pdftex]{hyperref}
\usepackage{geometry}
\DeclareMathOperator{\sech}{sech}

\title{Investigating overtaking collisions of solitary waves in the Schamel equation}
\author{Marcelo V. Flamarion$^{1}$, Efim Pelinovsky$^{2,3}$ and Ekaterina Didenkulova$^{2, 3}$}
\date{}

\begin{document}
\maketitle
\begin{center}
{\footnotesize $^1$Unidade Acad{\^ e}mica do Cabo de Santo Agostinho, \\
UFRPE/Rural Federal University of Pernambuco, BR 101 Sul, Cabo de Santo Agostinho-PE, Brazil,  54503-900 \\
marcelo.flamarion@ufrpe.br }

\vspace{0.3cm}
{\footnotesize $^{2}$Institute of Applied Physics, 46 Uljanov Str., Nizhny Novgorod 603155, Russia. \\
$^{3}$Faculty of Informatics, Mathematics and Computer Science, HSE University, Nizhny Novgorod 603155, Russia. }




\end{center}


\begin{abstract} 
This article presents a numerical investigation of overtaking collisions between two solitary waves in the context of the Schamel equation. Our study reveals different regimes characterized by the behavior of the wave interactions. In certain regimes, the collisions maintain two well-separated crests consistently over time, while in other regimes, the number of local maxima undergoes variations following the patterns of $2\rightarrow 1\rightarrow 2\rightarrow 1\rightarrow 2$ or $2\rightarrow 1\rightarrow 2$. These findings demonstrate that the geometric Lax-categorization observed in the Korteweg-de Vries equation (KdV) for two-soliton collisions remains applicable to the Schamel equation. However, in contrast to the KdV, we demonstrate that an algebraic Lax-categorization based on the ratio of the initial solitary wave amplitudes is not feasible for the Schamel equation. Additionally, we show that the statistical moments for two-solitary wave collisions are qualitatively similar to the KdV equation and the phase shifts after soliton interactions are close to ones in integrable KdV and modified KdV models.

	\end{abstract}

\section{Introduction}

Solitons, also referred to as solitary waves, are localized waveforms that maintain their shape and travel at a constant speed. They have applications in diverse fields, including water wave modeling for tsunamis, signal propagation in neuroscience and optical fibers, biological studies of proteins and DNA, localized magnetization in magnets, and nuclear physics \cite{Baines, Joseph}. In particular, the interaction of soliton with external force fields has been extensively investigated in the past decades \cite{Ermakov, Flamarion-Pelinovsky:2022a, Chaos:FP, Grimshaw94, Grimshaw96, Grimshaw:1993, Grimshaw2002, Kim, Lee, LeeWhang, Pelinovsky:2002, Whitham2, Wu1}. The dynamics and interactions of solitons are often described by the Korteweg-de Vries equation (KdV), a widely used equation. Zabusky and Kruskal \cite{Zabusky} conducted pioneering numerical investigations on the collisions of two solitons in the KdV equation. Their findings revealed that the solitons return to their initial forms after the collision. 

Lax \cite{Lax} later made significant contributions by categorizing two-soliton interactions in the KdV equation based on geometric and algebraic considerations. This classification, known as the Lax categorization, solely depends on the ratio of the initial soliton amplitudes and determines the number of local maxima during the collision. This categorization has proven valuable in explaining the dynamics of soliton interactions and has been supported by experimental studies and numerical simulations. Experimental studies by Weidman and Maxworthy \cite{Maxworthy} in a water-filled tank and numerical simulations by Mirie and Su \cite{Su} further validated the Lax categorization for soliton collisions. Their work confirmed that the observed soliton interactions aligned with Lax  geometric classification. Additionally, Craig et al. \cite{Craig:2006} investigated two-soliton interactions experimentally and numerically in the context of the Euler equations. Their findings supported Lax geometric categorization, although the algebraic categorization exhibited a different range from Lax predictions. Along the same lines, Flamarion \cite{COAM:2022} showed numerically that the non-integrable Whitham equation holds the Lax geometric classification, however a Lax-algebraic classification based on the amplitude of the two initial solitary waves is not possible. Similar results were also confirmed by Flamarion and Ribeiro-Jr \cite{Collisions} in the context of the forced Korteweg-de Vries equation. Analogous  studies have also been conducted in the Serre equations framework \cite{Antonopoulos:2017, Dutykh:2018} and in a viscous core annular flow model \cite{Lowman:2014}

The Schamel equation, widely used in the field of plasma physics, serves as a mathematical model for studying nonlinear wave phenomena, particularly the dynamics of ion acoustic waves in plasmas \cite{Schamel:1972, Schamel:1973, Ali:2017, Chowdhury:2018, Mushtaq:2006, Williams:2014, Saha:2015a, Saha:2015b}. Initially proposed by Schamel \cite{Schamel:1973}, the equation incorporates the interactions between ions and electrons in a plasma, accounting for collisions, plasma density, and temperature gradients. It offers valuable insights into various phenomena, including solitons. Solitons governed by the Schamel equation exhibit intriguing characteristics like self-trapping and particle acceleration.  The distinctive feature of the Schamel equation, distinguishing it from the well-known KdV equation, lies in its nonlinear representation, with the modular term playing a significant role. This characteristic renders the Schamel equation non-integrable and introduces mathematical challenges due to the non-analytic nature of the associated function.

In this study, we conduct a numerical investigation on the overtaking collisions of two solitary waves using the Schamel equation. Our findings confirm the validity of the three geometric categories as described by Lax \cite{Lax}. However, we demonstrate that a similar algebraic categorization based on the ratio of the amplitude of the initial solitary waves, as proposed by Lax, is not applicable to the Schamel equation. Furthermore, the statistical moments of the two-solitary wave collisions are computed and compared qualitatively with the KdV equation.

This article is organized as follow. In section 2 we present the Schamel equation.  The results are presented in sections 4, 5 and 6.  The conclusion is presented in section 7.

\section{The Schamel equation}
In our research, we investigate solitary wave interactions by focusing on the Schamel equation in its canonical form
\begin{equation}\label{Schamel1}
u_{t} +\sqrt{|u|}u_{x}+u_{xxx}=0.
\end{equation}
Within this equation, the variable $u$ represents the wave field at a specific position $x$ and time $t$. It is worth noting that the Schamel equation is a Hamiltonian equation, meaning it possesses a Hamiltonian function that governs its behavior. The Hamiltonian associated with this equation is defined as follows
\begin{equation}\label{Hamiltonian}
\mathcal{H} = \int_{-\infty}^{+\infty}\Big[-\frac{1}{2}u_{x}^{2}+\frac{4}{15}\mathrm{sign}(u)|u|^{5/2}\Big] dx.
\end{equation}
By expressing Equation (\ref{Schamel1}) in Hamiltonian form with respect to the functional $\mathcal{H}$, we can establish a relationship between the wave dynamics and the Hamiltonian. This connection is expressed by the following equation
\begin{equation*}
u_{t} = \frac{\partial}{\partial x}\Bigg[\frac{\delta\mathcal{H}}{\delta u}\Bigg],
\end{equation*}
where the functional derivative of the Hamiltonian with respect to $u$ is given by
\begin{equation*}
\frac{\delta\mathcal{H}}{\delta u}=u_{xx}+\frac{2}{3}\mathrm{sign}(u)|u|^{3/2}.
\end{equation*}

One intriguing feature of the Schamel equation is the invariance of its Hamiltonian $\mathcal{H}$ due to the absence of explicit time dependence. This invariance implies that the Hamiltonian remains constant throughout the evolution of the wave system. Furthermore, the Schamel equation (\ref{Schamel1}) possesses an additional invariant, known as the Casimir invariant or the mass invariant. This quantity is defined by the following integral:
\begin{equation}\label{mass}
M(t) = \int_{-\infty}^{+\infty}u(x,t) dx,
\end{equation}
and it characterizes the mass or the total "amount" of the wave field at any given time $t$. In addition to the mass invariant, the equation also exhibits a momentum invariant, given by
\begin{equation}\label{momentum}
P(t) = \int_{-\infty}^{+\infty}u^{2}(x,t) dx.
\end{equation}
These invariants, namely the Hamiltonian (\ref{Hamiltonian}), the mass (\ref{mass}), and the momentum (\ref{momentum}), play a crucial role in evaluating the accuracy and reliability of numerical methods employed to solve the Schamel equation (\ref{Schamel1}).

The Schamel equation (\ref{Schamel1}) supports solitary waves as its solutions. These solitary waves can be described by the following expressions
\begin{equation}\label{solitary}
u(x,t) = a\sech^{4}\left(k(x-ct)\right), \mbox{ where }  c = \frac{8\sqrt{|a|}}{15} \mbox{ and } k = \sqrt{\frac{c}{16}}.
\end{equation}
In this context, $a$ represents the amplitude of the solitary wave, which can be positive or negative. The parameter $c$ denotes the speed of the solitary wave and $k$ characterizes its wavenumber.

\section{Numerical methods}
\begin{figure}[h!]
	\centering	
	\includegraphics[scale =0.99]{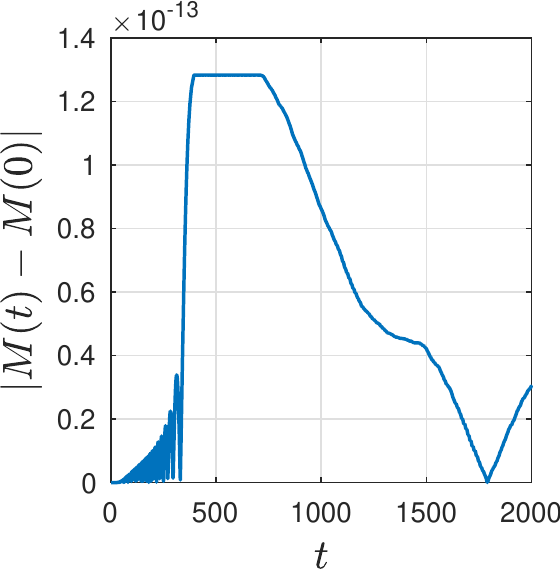}
	\includegraphics[scale =0.99]{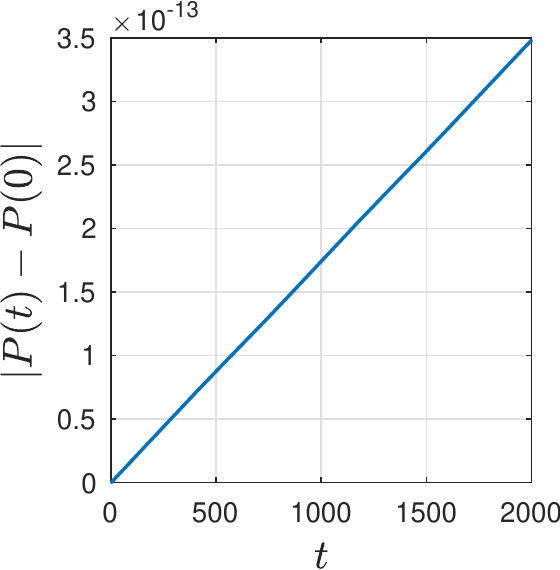}
	\includegraphics[scale =0.99]{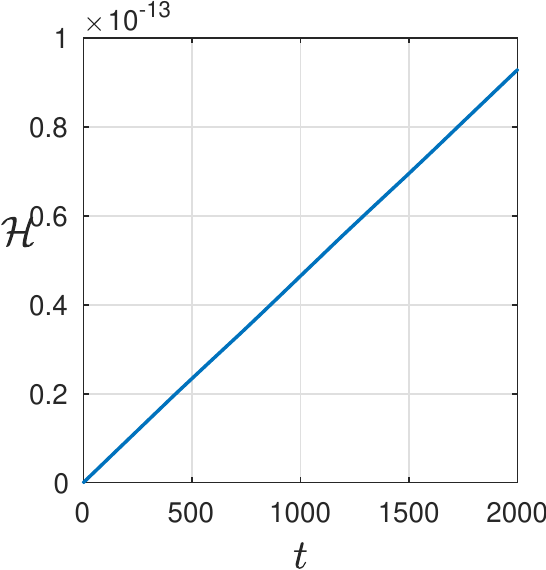}
	\caption{ The conserved quantities for the Schamel equation (\ref{Schamel1}) with a solitary wave with amplitude $a=1$. From left to right, the mass conservation, momentum conservation and the Hamiltonian. }
	\label{FigConserved}
\end{figure}

A Fourier pseudospectral method combined with an integrating factor is utilized to numerically solve the Schamel equation (\ref{Schamel1}). The computational domain chosen for the simulation is a periodic interval $[-L, L]$, discretized with an equidistant grid consisting of $N$ points. This grid configuration facilitates precise approximation of spatial derivatives, as discussed in \cite{Trefethen:2000}. To mitigate the influence of spatial periodicity, a sufficiently large computational domain is employed. For the temporal evolution of the equation, the classical fourth-order Runge-Kutta method is employed, employing discrete time steps of size $\Delta t$. Typical simulations employ parameter values such as $L=200$, $N=2^{13}$, and $\Delta t =0.005$. Notably, it has been verified that the numerical method preserves the quantities of mass, momentum, and the Hamiltonian. Figure \ref{FigConserved} depicts the preservation of these quantities over time for an initial solitary wave with an amplitude of $a=1$.

\section{Solitary wave interactions}

We investigate the intriguing phenomenon of overtaking collisions between two solitary waves in the context of the Schamel equation. Interactions of two solitons were studied previously for various models as a fundamental process of nonlinear wave dynamics, including integrable KdV-like models \cite{Anco:2011, Collisions, Pelinovsky:2015, Pelinovsky:2013a, Shurgalina:2018a, Shurgalina:2018b, Slyunyaev:1999} and non-integrable models \cite{COAM:2022, Craig:2006, Kachulin:2018}. Such processes are especially significant for multi-soliton dynamics known as soliton/breather turbulence \cite{Didenkulova:2022, Shurgalina:2016, Shurgalina:2017b}. To explore this phenomenon, we consider two solitary waves denoted as $S_1$ and $S_2$, with respective amplitudes $A_1$ and $A_2$. In our setup, we ensure that the amplitude of $S_1$ is greater than that of $S_2$ (as indicated in equation (\ref{solitary})), setting the stage for an interesting interaction. Initially, we position these solitary waves at a considerable distance from each other, ensuring that there is a significant separation between their respective crests. This configuration allows for clear observation and analysis of the overtaking collision dynamics that will unfold between $S_1$ and $S_2$. By examining this specific scenario, we aim to gain deeper insights into the behavior and characteristics of solitary waves as they interact and influence each other To achieve this, the initial data is taken as
\begin{equation}\label{initial}
u(x,0) = S_{1}(x+20)+S_{2}(x-20).
\end{equation}

It is worth noting that Lax classified the characteristics of two-solitary wave interactions in the KdV equation based on the ratio of the initial amplitudes $A_1$ and $A_2$ as follows
\begin{itemize}
	\item[{\bf (A)}] For $A_{1}/A_{2}<(3+\sqrt{5})/2\approx 2.62$, the solution of the KdV equation exhibits two distinct and separate crests, indicating the presence of two local maxima at any given time $t$.

	\item[{\bf(C)}] When $A_{1}/A_{2}>3$, the number of local maxima undergoes a transition $2\rightarrow 1\rightarrow 2$ during the interaction. This transition implies that the solitary waves merge temporarily, forming a wave with a single local maximum for a certain period of time.
	
	\item[{\bf (B)}]  	This case exhibits characteristics of both cases {\bf(A)} and {\bf(C)}. Specifically, the interaction between the two solitary waves can be described in the following steps: (i) Initially, the solitary waves are well-separated, resulting in two distinct crests. (ii) Over time, the solitary waves merge and form a wave with a single local maximum. (iii) The wave subsequently splits into two, giving rise to two local maxima. (iv) The two waves then combine again to form a single maximum. (v) Finally, the crests separate, and the two waves with two crests reappear at later times. Mathematically, we observe that during the collision, the number of local maxima follows a pattern of $2\rightarrow 1\rightarrow 2\rightarrow 1\rightarrow 2$. This case occurs when $(3+\sqrt{5})/2<A_{1}/A_{2}<3$.
\end{itemize}
After the collision, the solitary waves undergo a phase shift, meaning their crests are slightly displaced from the paths of the incoming centers.

The following graphs are presented in the reference frame of $S_2$, where $S_1$ moves from left to right and intersects with the stationary solitary wave $S_2$. It is important to note that the Schamel equation (\ref{Schamel1}) is not integrable. Nevertheless, the two solitary waves exhibit nearly identical shapes following their collision. However, it should be emphasized that this is not an exact two-solitary wave solution due to the presence of a small dispersive radiation that propagates to the left. Details regarding the amplitude of the solitary waves after the interaction are describe in Table \ref{table0} and the maximum value  between the two solitary waves.
\begin{table}[h!]
\centering
\begin{tabular}{c|c|c|c|c|c|c|c|c}\hline\hline
\multicolumn{2}{c|}{Initial Amplitudes}    & \multicolumn{2}{c|}{Amplitudes after the interaction} & \multicolumn{1}{c|}{Maximum value between two crests}  & \multicolumn{1}{c|}{Category}  \\   \hline

$A_{1}$  & $A_{2}$  & $A_{1}$ & $A_{2}$ & $A_{max}$ &     \\   \hline
$1.000$  & $0.800$  & $1.000$ & $ 0.799$  & $0.194$   & \bf{A}  \\   \hline              
$1.000$  & $0.300$  & $ 1.000$ & $0.298$ & $ 0.684$    & \bf{C}   \\   \hline 
$1.000$  & $0.350$  & $1.000$ & $0.348$  & $ 0.635$ & \bf{B} \\   \hline     
\end{tabular}
\caption{Amplitude of the solitary waves at different stages of the interactions. The amplitudes after the interaction were computed at large times so that the shapes of $S_1$ and $S_2$ are preserved.}\label{table0}
\end{table}

{ The KdV equation provides a prediction that the maximum value attained between two solitary waves is equal to the difference between the amplitudes of the initial solitary waves. However, in the case of the Schamel equation, Table \ref{table0} demonstrates a slight deviation from this prediction. The observed value is slightly lower than what is anticipated by the KdV equation. This discrepancy arises due to the loss of mass experienced by the solitary wave $S_1$ as a result of dispersive effects during its interaction with $S_2$. The dispersive effects play a significant role in this phenomenon, causing a reduction in the amplitude of $S_1$ after the interaction. This behavior is consistently observed across all simulations, irrespective of the specific type of interaction under consideration. These findings underscore the impact of dispersive processes on the dynamics of solitary waves and highlight the importance of accounting for such effects in the study of wave interactions.}



\begin{figure}[h!]
	\centering	
	\includegraphics[scale =1]{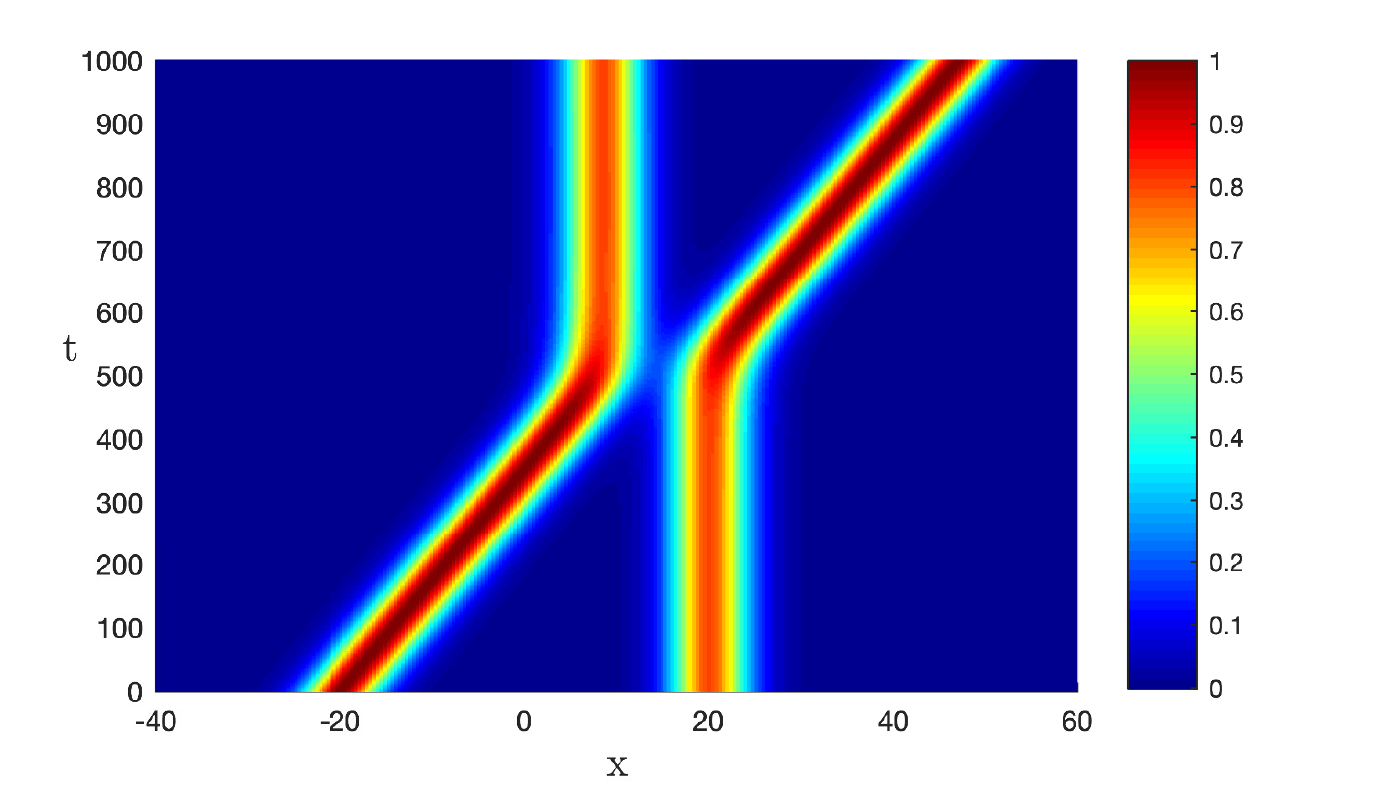}
	\includegraphics[scale =1]{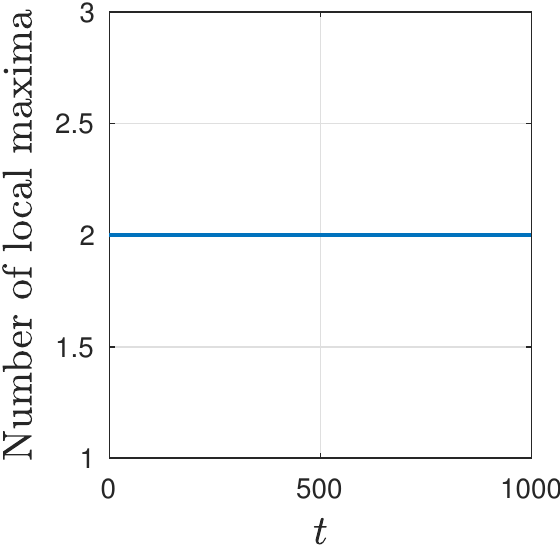}
	\includegraphics[scale =1]{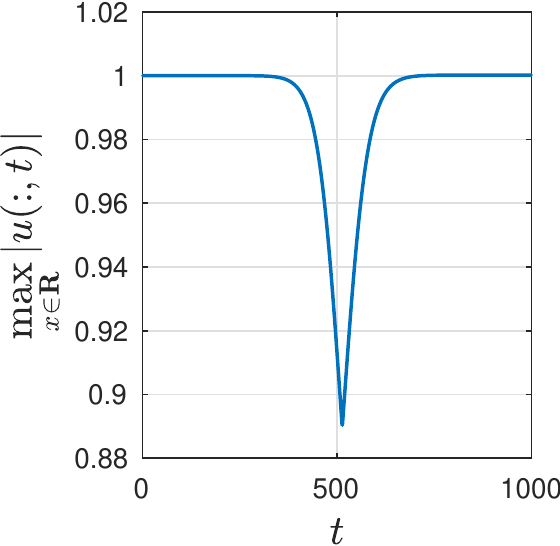}
	\caption{Top: Overtaking collision of two solitary waves for the Schamel equation -- category {\bf(A)}. Bottom: (Left) The number of local maxima as a function time. Right: The maximum amplitude as a function of time. Parameters $A_{1}=1.000$, $A_{2}=0.800$.} 
	\label{colisaoA}
\end{figure}
\begin{figure}[h!]
	\centering	
	\includegraphics[scale =1.05]{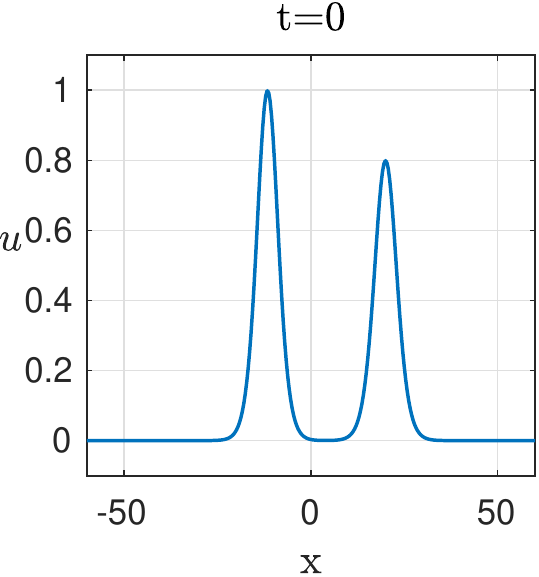}
	\includegraphics[scale =1.05]{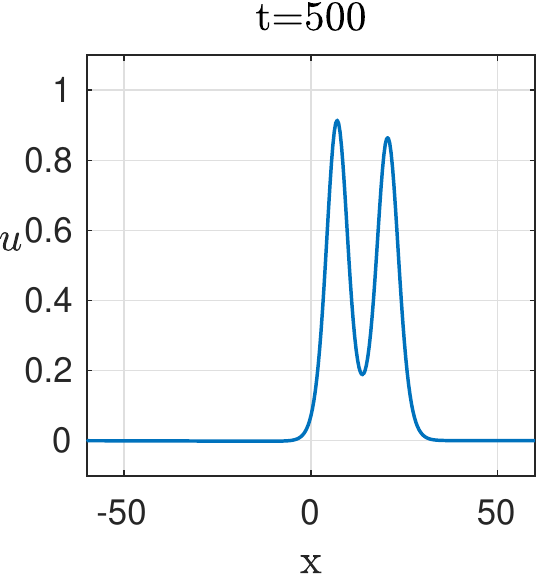}
        \includegraphics[scale =1.05]{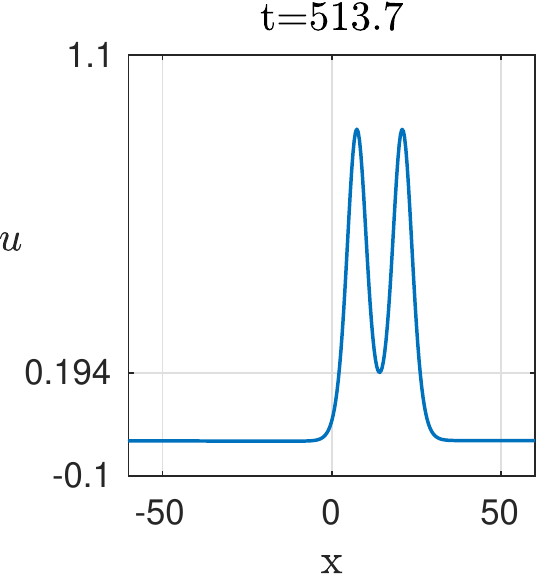}
	\includegraphics[scale =1.05]{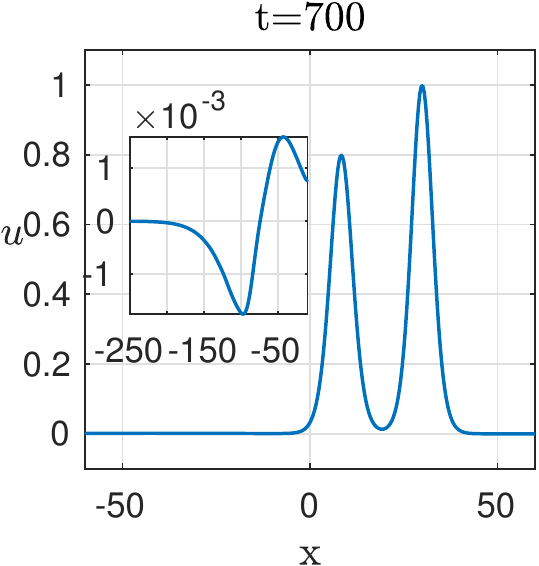}
	\caption{A series of snapshots of Figure \ref{colisaoA}. The last snapshot displays the dispersion radiation produce during the solitary wave collision.} 
	\label{scolisaoA}
\end{figure}

Figure \ref{colisaoA} (top) illustrates the collision between two solitary waves. Initially, the solitary waves are separated, and over time,  $S_1$ propagates towards  $S_2$, leading to a collision. During this collision, the solitary wave $S_1$ starts to diminish while soliton $S_2$ begins to grow until the two waves exchange their roles (refer to Figure \ref{colisaoA} (bottom-right)) { and a phase lag is observed.} More details of this interaction can be seen in the series of the snapshots in Figure \ref{scolisaoA}. Notice that a small dispersive tails appears during the collision of the solitary waves. This occurs because the Schamel equation is not an integrable equation. It's worth noting that at any given time, there exist two local maxima, indicating that the wave crests of the two solitary waves never intersect (as depicted in Figure \ref{colisaoA} (bottom-left)) and the maximum amplitude during the interaction is the same as the amplitude of $S_1$ (see Figure \ref{colisaoA} (bottom-right))  and Table \ref{table0}. {  This observation implies that the solitary wave $S_1$ remains unchanged despite the presence of dispersive effects. Moreover, the amplitude of the interacting solitary wave $S_2$ experiences minimal alteration. As a result, this collision can be characterized as highly elastic, with the waves exhibiting a remarkable ability to retain their fundamental characteristics during the interaction.} The particular scenario corresponds to case {\bf (A)} according to Lax categorization.

To elaborate further on this discussion, let's delve into the dynamics of the solitary wave collision. Initially, when the solitary waves are sufficiently far apart, they travel undisturbed, maintaining their individual shapes and amplitudes. As time progresses,  $S_1$ accelerates towards  $S_2$ due to their mutual attraction. Eventually, they come close enough to interact, resulting in the collision. During the collision phase, an interesting phenomena unfold. The solitary wave $S_1$, being the larger one, undergoes a reduction in size, while $S_2$ experiences an increase in size. This exchange of energy between the solitary waves lead to a temporary merging of their respective wave profiles. However, due to the intrinsic nature of the solitary waves, the wave crests remain distinct, ensuring that they never coincide.
\begin{figure}[h!]
	\centering	
	\includegraphics[scale =1]{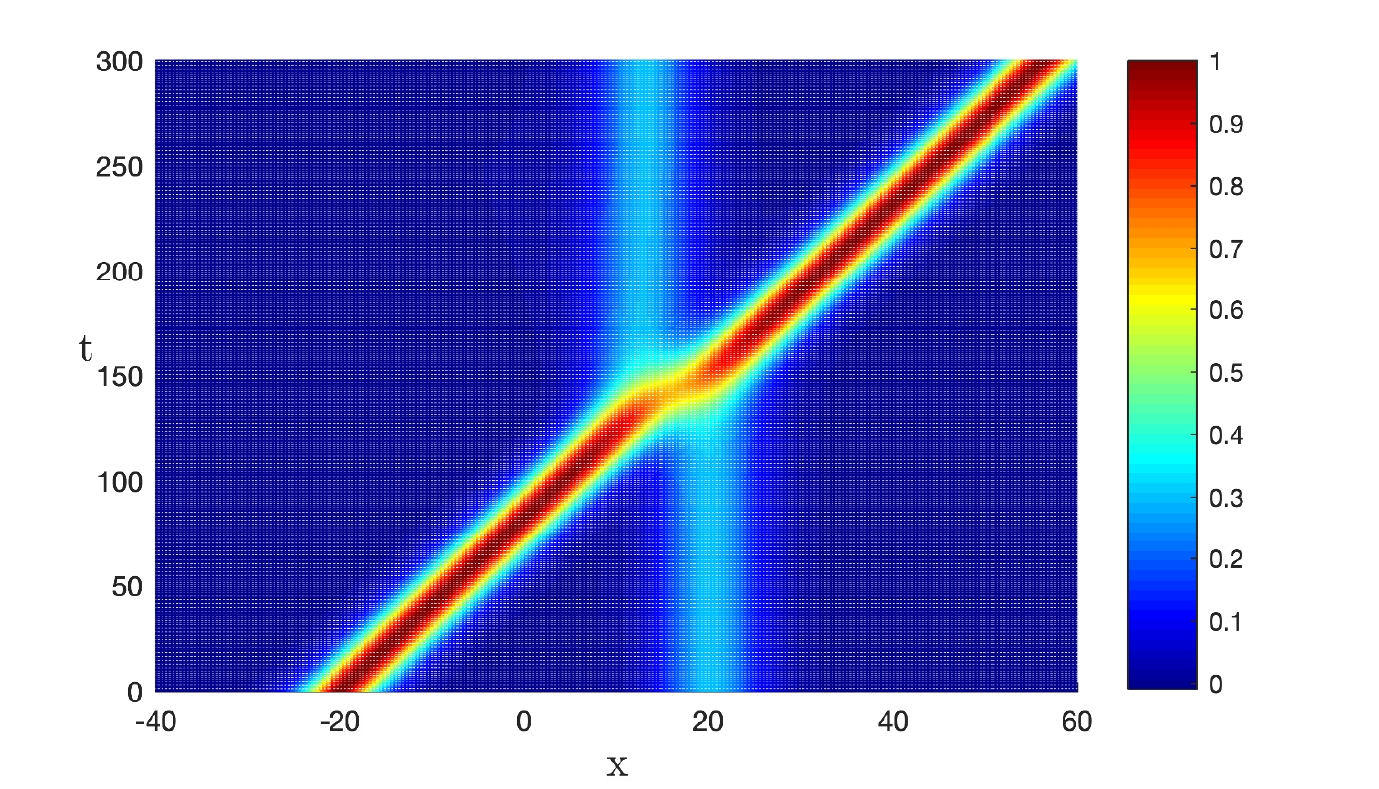}
	\includegraphics[scale =1]{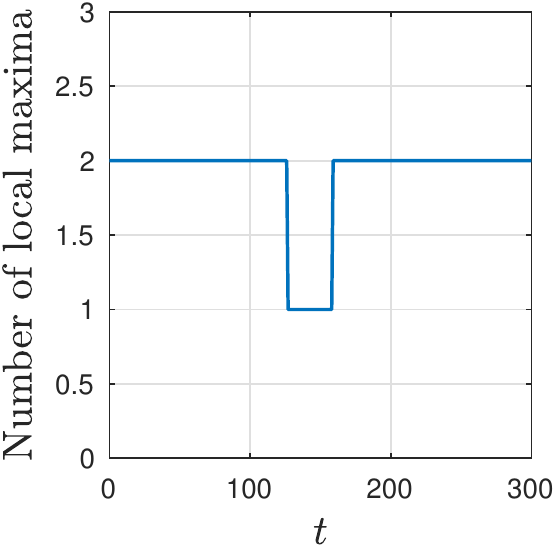}
	\includegraphics[scale =1]{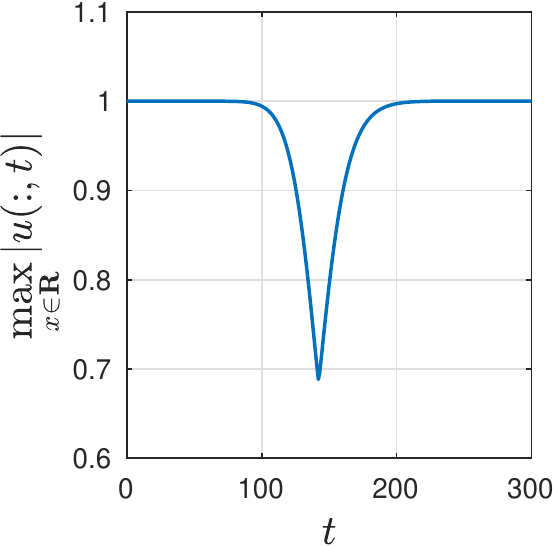}
	\caption{Top: Overtaking collision of two solitary waves for the Schamel equation -- category {\bf(C)}. Bottom: (Left) The number of local maxima as a function of time. Right: The maximum amplitude as a function of time.  Parameters $A_{1}=1.000$, $A_{2}=0.300$.} 
	\label{colisaoC}
\end{figure}
\begin{figure}[h!]
	\centering	
	\includegraphics[scale =1.05]{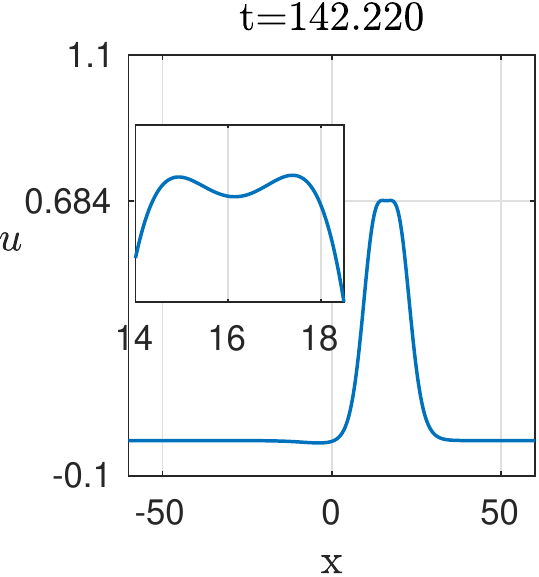}
	\includegraphics[scale =1.05]{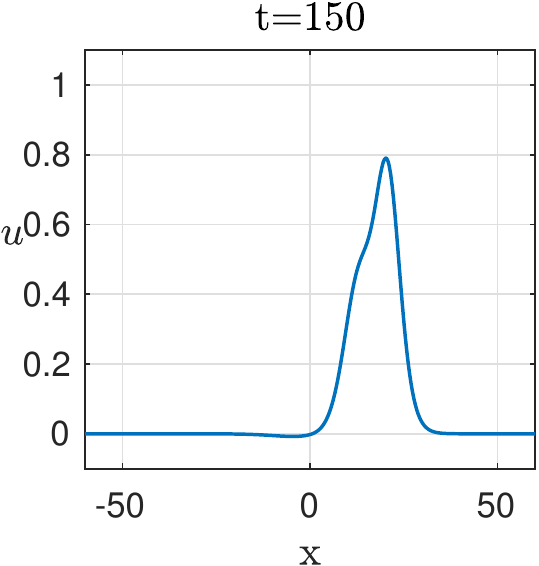}
        \includegraphics[scale =1.05]{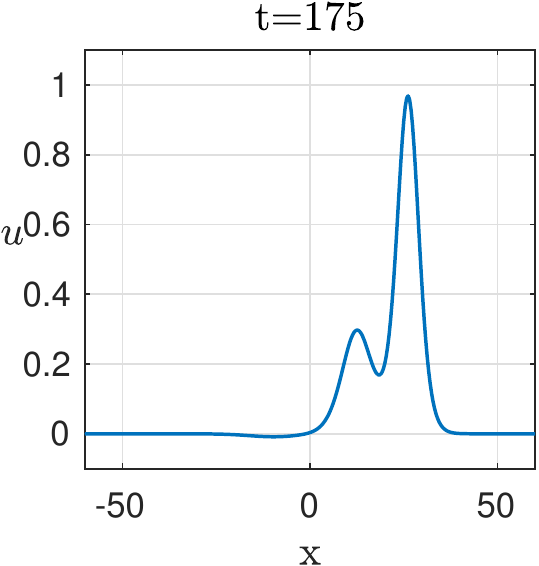}
	\includegraphics[scale =1.05]{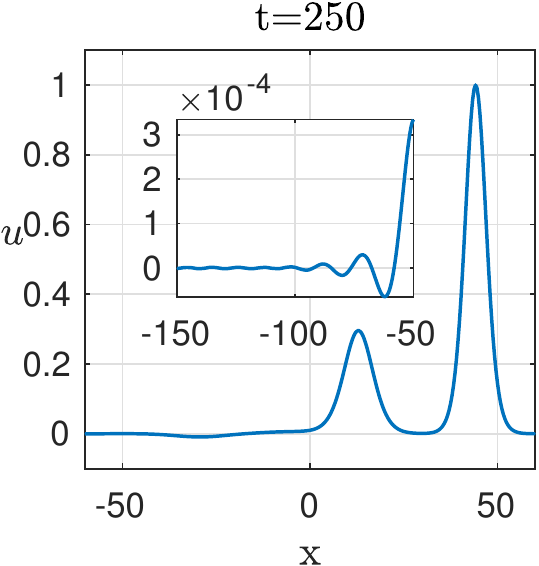}
	\caption{A series of snapshots of Figure \ref{colisaoC}. The last snapshot displays the dispersion radiation produce during the solitary wave collision.} 
	\label{scolisaoC}
\end{figure}

In Figure \ref{colisaoC} (top), we observe the collision between two solitary waves, where the solitary wave $S_1$ is significantly larger than  $S_2$ (with amplitudes $A_1 = 1.000$ and $A_2 = 0.300$, respectively). This particular case demonstrates a  phenomenon in which the two solitary waves merge to form a single local maximum for a certain period of time (see Figure \ref{colisaoC} (bottom-left)). Subsequently,  $S_1$ absorbs  $S_2$, and later on, reemerges with a phase lag in the trajectories of the crest (as shown in Figure \ref{colisaoC} (top)). During the collision phase, a captivating phenomenon occurs. The solitary wave $S_1$, being larger in amplitude, acts as a dominant solitary wave and engulfs  $S_2$. As a result, the two solitary waves merge and combine to form a single localized maximum, effectively occupying the same space temporarily. This merging of the solitary waves reflects the strong interaction between them and showcases the nonlinear nature of the wave system under consideration. Additional information regarding this interaction is observable in the sequence of images depicted in Figure \ref{scolisaoC}. It is worth noting that a diminutive dispersive tail  emerges immediately following the collision of the solitary waves. Notice that the maximum amplitude during  interaction is the same as the amplitude of the larger solitary wave, see Figure \ref{colisaoC} (bottom-right) and Table \ref{table0}.  In the context of wave interactions, it is observed that the solitary wave $S_2$ remains unaffected and retains its integrity after the interaction takes place. However, the solitary wave $S_1$ experiences a slight decrease in amplitude due to the nonintegrability inherent in the Schamel equation. It is important to note that this reduction in amplitude is relatively minor when compared to the original amplitude of $S_2$. Therefore, we can conclude that the collision between the two waves can be considered as nearly elastic in nature. This finding highlights the remarkable resilience of solitary waves, such as $S_2$, which manage to preserve their structure even during interactions. Although $S_1$ undergoes a small amplitude loss, it is overshadowed by the robustness of $S_2$. This behavior indicates that the system possesses a certain level of energy conservation, with most of the initial energy being transferred to $S_2$ while only a fraction is dissipated through the non-integrable dynamics of the Schamel equation. The nearly elastic collision between these solitary waves showcases the intriguing dynamics and intricate interplay between nonintegrability and wave behavior. Further investigations into the properties and behaviors of solitary waves can shed light on the underlying physics and offer insights into the fundamental principles governing wave interactions in non-integrable systems

\begin{figure}[h!]
	\centering	
	\includegraphics[scale =1]{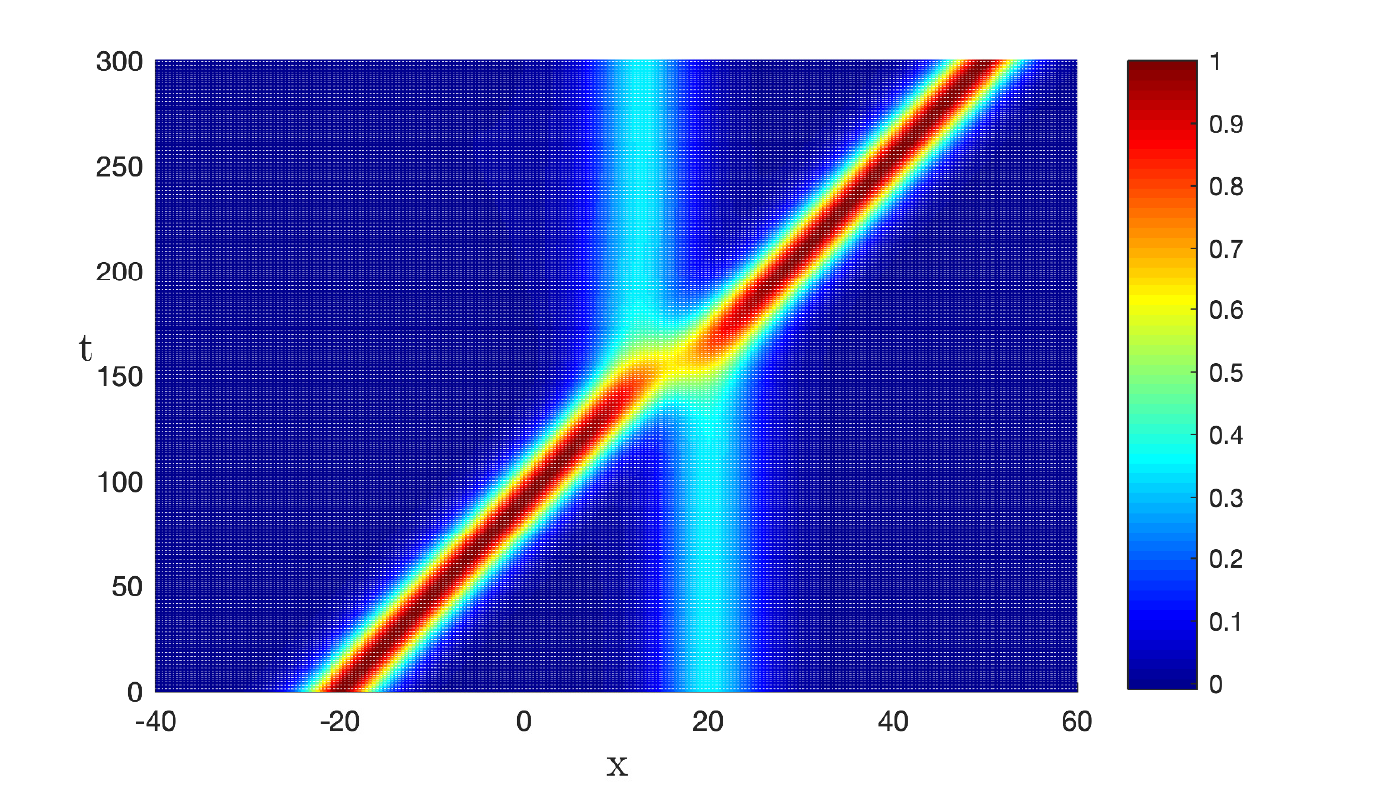}
	\includegraphics[scale =1]{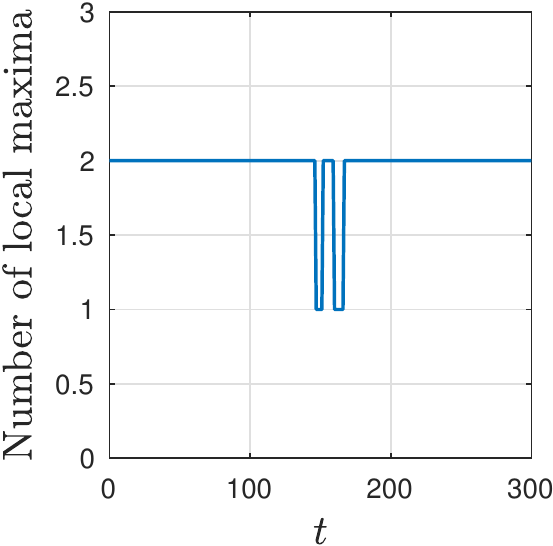}
	\includegraphics[scale =1]{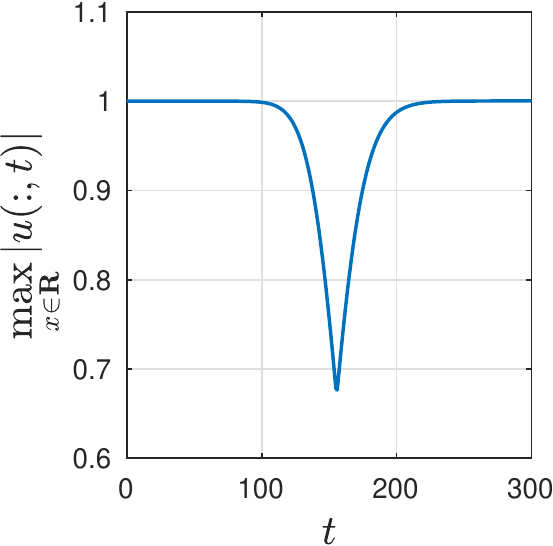}
	\caption{Top: Overtaking collision of two solitary waves for the Schamel equation -- category {\bf(B)}. Bottom: (Left) The number of local maxima as a function of time. Right: The maximum amplitude as a function of time. Parameters $A_{1}=1.000$, $A_{2}=0.350$.} 
	\label{colisaoB}
\end{figure}
\begin{figure}[h!]
	\centering	
	\includegraphics[scale =1.05]{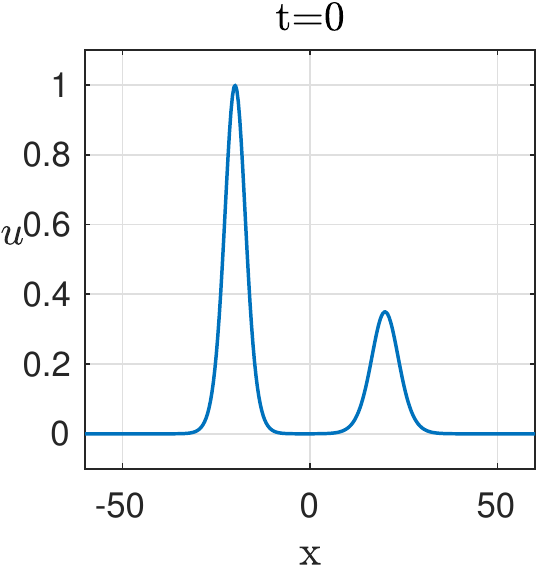}
	\includegraphics[scale =1.05]{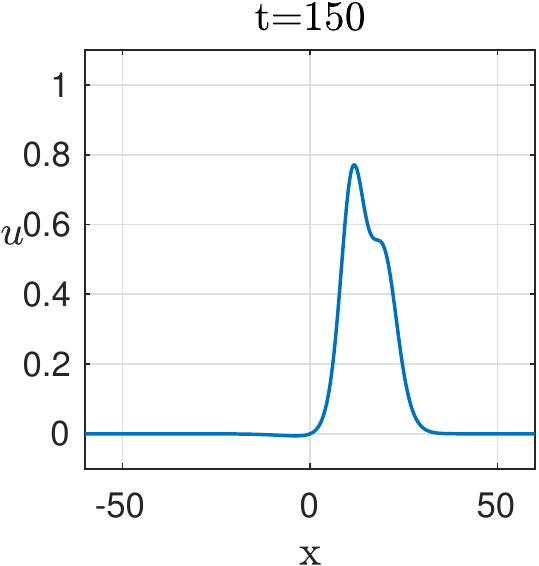}
        \includegraphics[scale =1.05]{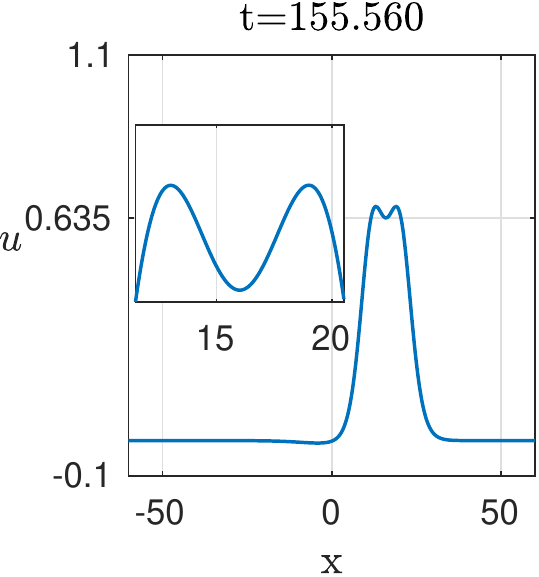}
	\includegraphics[scale =1.05]{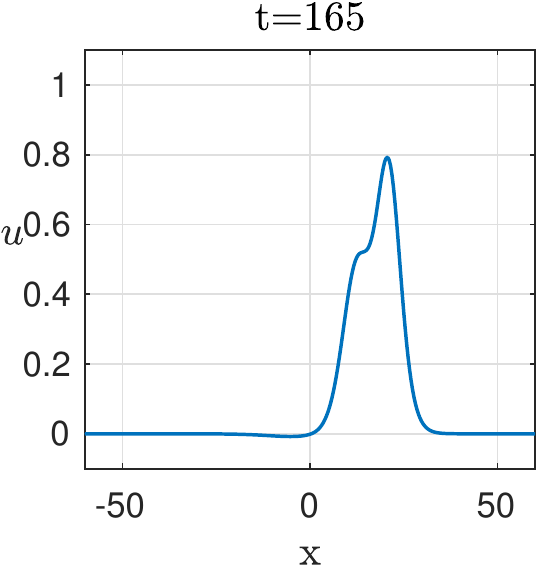}
	\includegraphics[scale =1.05]{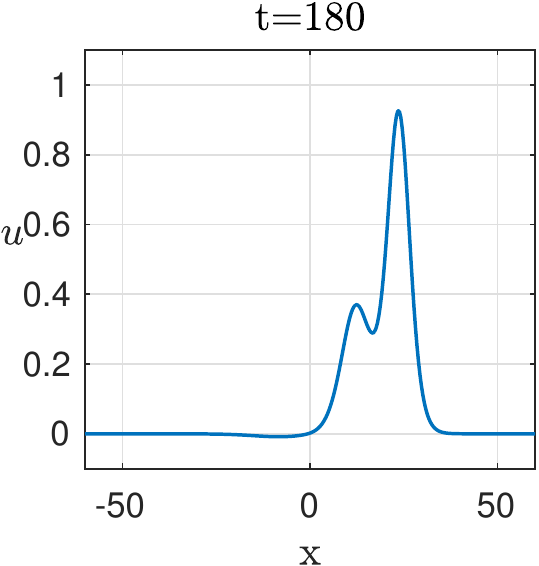}
	\includegraphics[scale =1.05]{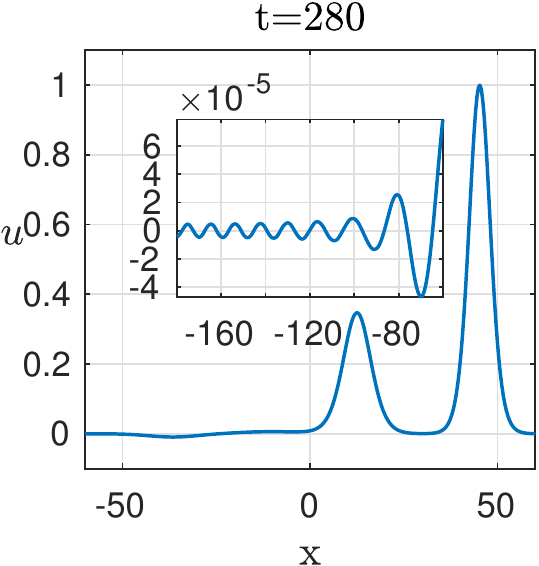}
	\caption{A series of snapshots of Figure \ref{colisaoB}. The last snapshot displays the dispersion radiation produce during the solitary wave collision.} 
	\label{scolisaoB}
\end{figure}

Lastly, let us examine a more complex interaction between two solitary waves, as depicted in Figure \ref{colisaoB}. In this scenario, we have initial  solitary waves with amplitudes $A_1 = 1.000$ and $A_2 = 0.350$. During their interaction, the solitary wave $S_1$ engulfs  $S_2$, resulting in the formation of a single local maximum. Subsequently, the solitary wave $S_2$ splits into two distinct waves, followed by the observation of a single local maximum once again. As time progresses, the waves gradually move apart, eventually revealing two waves with well-defined crests. Details of the variation of the number of local maxima are described in Figure \ref{colisaoB} (bottom-left) and variations of the amplitude during the interactions in Figure \ref{colisaoB} (bottom-right). Further insight into this interaction can be gleaned from the series of snapshots presented in Figure \ref{scolisaoB}. It is noteworthy that a subtle dispersive tail manifests itself shortly after the collision of the solitary waves. { Furthermore, it is noteworthy that the amplitude of $S_2$ remains remarkably preserved even after the interaction between the two solitary waves. This observation serves to further reinforce the notion that collisions involving solitary waves exhibit a high degree of elasticity. }

{}

While it is feasible to categorize the two-soliton interaction based on the amplitude ratio of the initial solitons for the KdV equation \cite{Lax} and the full Euler equations \cite{Craig:2006}, such classification is not applicable to the Schamel equation. Table \ref{table2} indicates that the region where collision of type {\bf (B)} occurs is thin. Thus, we consider more collisions with amplitude ratios close to $2.800$. Consequently, we conclude that the  algebraic Lax categorization fails for the Schamel equation. Table \ref{table2} provides examples illustrating that an algebraic categorization, following the Lax approach, using the ratio of initial solitary wave amplitudes is not viable for the Schamel equation.
\begin{table}[h!]
\centering
\begin{tabular}{c|c|c|c|c|c|c}\hline\hline
\multicolumn{2}{c|}{Amplitudes} & \multicolumn{1}{c|}{Ratio}   &  \multicolumn{1}{c}{Category of Schamel equation}  \\   \hline
$A_{1}$ & $A_{2}$ & $A_1/A_2$ &  \\   \hline
1.000	& 0.950	& 1.053 	& \bf{A}  \\ \hline
1.000	& 0.900	& 1.111 	& \bf{A}  \\ \hline
1.000	& 0.850 	& 1.176 	& \bf{A}  \\ \hline
1.000	& 0.800	& 1.250 	& \bf{A}  \\ \hline
1.000	& 0.750 	& 1.333 	& \bf{A}  \\ \hline
1.000	& 0.700 	& 1.428 	& \bf{A}  \\ \hline
1.000	& 0.650 	& 1.538 	& \bf{A}  \\ \hline
1.000	& 0.600 	& 1.666 	& \bf{A}  \\ \hline
1.000	& 0.550 	& 1.818	& \bf{A}  \\ \hline
1.000	& 0.500 	& 2.000 	& \bf{A}  \\ \hline
1.000	 & 0.450 	& 2.222 	& \bf{A}  \\ \hline
1.000	 & 0.400 	& 2.500 	& \bf{A}  \\ \hline
1.000	& 0.350	& 2.857 	& \bf{B}  \\ \hline
1.000	& 0.300	& 3.333 	& \bf{C}  \\ \hline
1.000	& 0.250 	& 4.000	& \bf{C}  \\ \hline
1.000	 & 0.200 	& 5.000 	& \bf{C}  \\ \hline
1.000	 & 0.150 	& 6.666 	& \bf{C}  \\ \hline
1.000	 & 0.100 	& 10.000 	& \bf{C}  \\ \hline
\end{tabular}
\caption{Classification of the collision for different values of $A_1$ and $A_2$.}\label{table1}
\end{table}

\begin{table}[h!]
\centering
\begin{tabular}{c|c|c|c|c|c|c}\hline\hline
\multicolumn{2}{c|}{Amplitudes} & \multicolumn{1}{c|}{Ratio}   &  \multicolumn{1}{c}{Category of Schamel equation}  \\   \hline
$A_{1}$ & $A_{2}$ & $A_1/A_2$ &  \\   \hline
1.000	& 0.375	& 2.666 	& \bf{A}  \\ \hline
1.000	& 0.356	& 2.809 	& \bf{C}  \\ \hline
1.000	& 0.355	& 2.817 	& \bf{C}  \\ \hline
1.000	& 0.350	& 2.857 	& \bf{B}  \\ \hline
\end{tabular}
\caption{Classification of the collision for different values of $A_1$ and $A_2$.}\label{table2}
\end{table}

\section{Phase shifts during the solitary wave collision}
The phenomenon of soliton shifts resulting from their collisions with each other has been extensively studied for various integrable equations, such as Korteweg-de Vries (KdV) and modified Korteweg-de Vries (mKdV) equations \cite{Zakharov:1971,Lamb:1980, Newell:1985}. Typically, the formulas for soliton shifts are expressed in terms of soliton amplitudes. However, when formulated in terms of soliton speeds (c), the formula for phase shifts remains the same for both KdV and mKdV equations \cite{Shurgalina:2017} 
\begin{equation}\label{phaseshift}
\Delta x_{1,2} = \pm\frac{2}{\sqrt{c_{1,2}}}\ln\Bigg(\frac{\sqrt{c_{1}}+\sqrt{c_{2}}}{\sqrt{c_{2}}-\sqrt{c_{1}}}\Bigg),
\end{equation}
where $\Delta x$ is positive for the larger soliton and negative for the smaller one.

The analytical explanation of soliton shifts stems from kinetic theory \cite{Zakharov:1971,Kamchatnov:2005}, which applies to all integrable systems and is based on three fundamental principles: (i) solitons remain conserved during interactions, (ii) soliton interactions occur pairwise, and (iii) elastic collisions between solitons result in a change in their phases.

Formally, no theory currently exists for the Schamel equation due to its non-integrability. However, we conducted a comparison between the analytical phase shifts (\ref{phaseshift}) and the numerically calculated shifts using $x-t$ diagrams in the case of three considered solitary wave interactions, as presented in Table \ref{TablePhase}. The negative phase shifts of the smallest solitary waves in both numerical and analytical calculations ranged from $1.5\%$ to $11\%$. This variation can be attributed to the low radiation exhibited by solitary waves with small amplitudes. For larger solitary waves, the difference becomes more significant, particularly when the amplitudes of the solitary waves differ considerably ($23\%$ in the case of interaction between solitary waves with amplitudes $A_1=1$ and $A_2=0.3$). However, when the interacting solitary waves have closer amplitudes ($A_1=1$ and $A_2=0.8$), the discrepancy between the approximate numerical shift and the analytical shift is only $8\%$ for the largest solitary waves. Thus, it can be concluded that as the amplitude of the smallest solitary waves decreases, the numerically calculated phase shifts align more closely with the analytical values. However, the agreement between analytical and numerical phase shifts diminishes for the largest solitary waves due to their extended tails. In general, we can conclude that formula (\ref{phaseshift}) for "elastic" interactions provides a reasonably accurate description of the "inelastic" results, primarily due to the smallness of solitary wave tails.
\begin{table}[h!]
\centering
\begin{tabular}{c|c|c|c|c|c|c|c|c}\hline\hline
\multicolumn{2}{c|}{Amplitudes}    & \multicolumn{1}{c|}{Phase shift ($A_1$)} & \multicolumn{1}{c|}{Numerical phase shift ($A_1$)}  & \multicolumn{1}{c|}{Phase shift ($A_2$)} & \multicolumn{1}{c|}{Numerical phase shift ($A_2$)}  \\   \hline

$S_{1}$  & $S_{2}$    & & &  &  \\   \hline
$1.00$  & $0.80$   & $ 9.80$  & $10.60$   & $-10.40$ & $-11.70$  \\   \hline              
$1.00$  & $0.30$   & $5.20$ & $4.00$    & $-7.00$ & $-7.10$   \\   \hline 
$1.00$  & $0.35$   & $5.60$  & $ 4.50$ & $-7.30$ & $-7.50$  \\   \hline     
\end{tabular}
\caption{The phase shift experienced by the solitary waves computed through formula (\ref{phaseshift}) and numerically.}\label{TablePhase}
\end{table}

\section{Integral characteristtics}
The dynamics of multisoliton solutions are primarily influenced by the pairwise interactions between the individual solitons, as supported by analytical findings \cite{Zakharov:1971, Zakharov:2009}. These interactions play a crucial role in shaping the behavior and characteristics of the wave field. To gain a deeper understanding of these interactions and their impact on the statistical moments of the wave field, we have focused on examining four specific integrals, corresponding to four statistical moments
\begin{equation}\label{momentums}
M_{n}(t)=\int_{-\infty}^{+\infty}u^{n}(x,t)dx, \mbox{ where $n=1,2,3,4.$}
\end{equation}

\begin{figure}[h!]
	\centering	
	\includegraphics[scale =0.99]{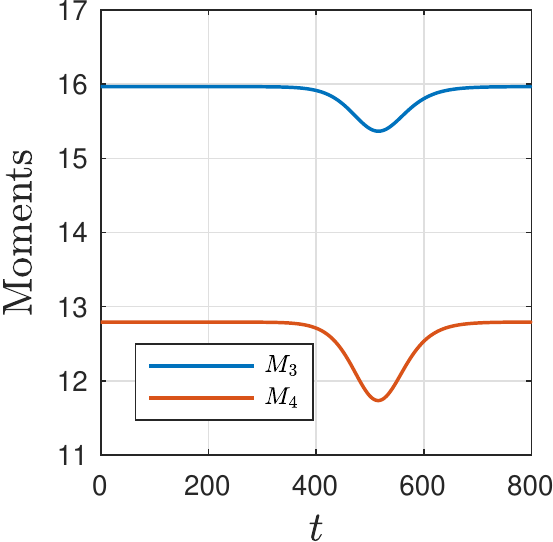}
	\includegraphics[scale =0.99]{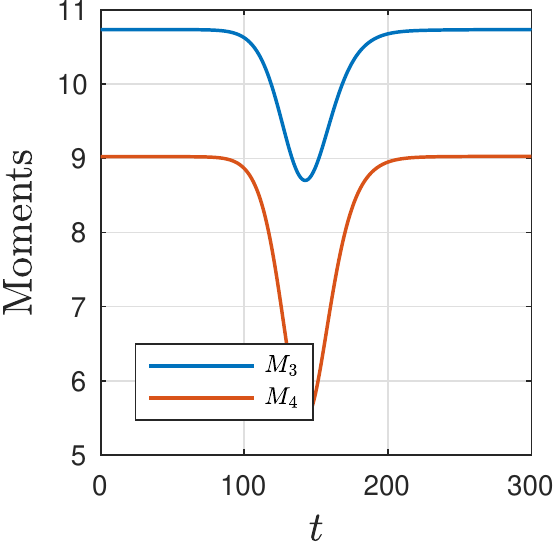}
	\includegraphics[scale =0.99]{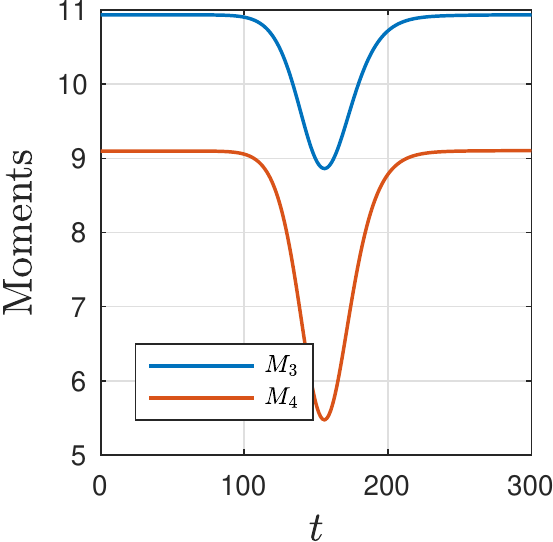}
	\caption{ Moments $M_3$ and $M_4$ as functions of time during interaction of the solitary waves of the Schamel equation for the collisions displayed in Figures \ref{colisaoA}, \ref{colisaoB} and \ref{colisaoC} respectively. }
	\label{FigMoments}
\end{figure}

The analysis of moments provides valuable insights into the statistical properties of the turbulent flow. While the first two integrals remain constant over time, representing the mean and variance of the system, the higher-order moments, such as the third and fourth moments, capture information about the skewness and kurtosis of the turbulence. Figure \ref{FigMoments} showcases the temporal evolution of the third and fourth moments during the interaction between two solitary waves governed by the Schamel equation. It is intriguing to note that these moments exhibit variation over time, unlike the first two moments. This temporal variation reflects the dynamic nature of the interaction process and the influence it has on the statistical properties of the system. Remarkably, the behavior of the third and fourth moments in the case of solitary wave interactions resembles the patterns observed in solitons governed by the KdV equation \cite{Pelinovsky:2015}. The reduction in both the third and fourth moments during the interaction phase suggests a decrease in the skewness and kurtosis of the resulting pulse. This phenomenon can be attributed to the corresponding reduction in the amplitude of the pulse as a consequence of the interaction between the solitary waves.

The analysis of moments provides a deeper understanding of the dynamics and statistical characteristics of turbulent systems. By examining the variations in higher-order moments, we gain insights into the nonlinear processes at play during the interaction of solitary waves. This knowledge contributes to the broader field of turbulence theory and aids in the development of more accurate models and predictions for complex fluid dynamics.

\section{Conclusion}
In this article, our numerical investigations have provided valuable insights into the interactions between two solitary waves within the framework of the Schamel equation. Through our study, we have confirmed the enduring validity of the Lax-geometric categorization for describing the interaction dynamics of the two solitary waves in this equation, akin to what has been observed in the well-known KdV equation. However, we have made a significant discovery that deviates from previous assumptions. We found that the algebraic categorization based on the ratio of the initial solitary wave amplitudes, which has been commonly employed in the KdV equation, is not applicable within the context of the Schamel equation. 
In addition, it was shown that some features of solitary wavess are similar to ones within integrable models due to small dispersion. Thus, moments for two-solitary wave collisions behave qualitatively similar to the KdV and mKdV equations. Phase shifts calculated for the KdV model can be used for approximate estimations in the Schamel equation.

Theses novels finding challenge our previous understanding and highlights the unique characteristics of solitary wave collisions in this particular equation. By unraveling the distinct behavior and properties of solitary wave collisions in the Schamel equation, our study contributes to the broader understanding of solitary wave dynamics. These findings open up new avenues for exploring the intricate dynamics and interactions of solitary waves in non-integrable systems.

\section{Acknowledgements}
M.V.F is grateful to IMPA for hosting him as visitor during the 2023 Post-Doctoral Summer Program. E.P. and E.D. are supported by Laboratory of Dynamical Systems and Applications NRU HSE, grant of the Ministry of science and higher education of the RF, ag. number 075-15-2022-1101.

	\section*{Declarations}
	
	\subsection*{Conflict of interest}
	The authors state that there is no conflict of interest. 
	\subsection*{Data availability}
	
	Data sharing is not applicable to this article as all parameters used in the numerical experiments are informed in this paper.

\end{document}